\def\ref{\par\noindent\hang}

\def\spose#1{\hbox to 0pt{#1\hss}}
\def\approxlt{\mathrel{\spose{\lower 3pt\hbox{$\sim$}}
        \raise 2.0pt\hbox{$<$}}}
\def\approxgt{\mathrel{\spose{\lower 3pt\hbox{$\sim$}}
        \raise 2.0pt\hbox{$>$}}}

\def\multleft#1{\hbox to size{\vbox {\halign {\lft{##}\cr #1}}\hfill}\par}
\def\multright#1{\hbox to size{\vbox {\halign {\rt{##}\cr #1}}\hfill}\par}

\def\degmark{^\circ}
\def\boxit#1{\vbox{\hrule\hbox{\vrule\kern3pt\vbox{\kern3pt
          #1 \kern3pt}\kern3pt\vrule}\hrule}}

\def\cm{{\rm\thinspace cm}}

\def\keV{{\rm\thinspace keV}}

\def\Msun{\hbox{$\rm\thinspace M_{\odot}$}}

\def\s{{\rm\thinspace s}}

\def\cm  {\hbox{$\cm\s^{-1}\,$}}

\documentstyle[emulateapj,psfig,times]{article}

\begin{document}
\title{Iron Line Reverberation Mapping With Constellation-X}
\author{A. J. Young}
\affil{Institute of Astronomy, Madingley Road, Cambridge CB3 0HA}
\and
\author{C. S. Reynolds\thanks{Hubble Fellow}}
\affil{JILA,University of Colorado, Campus Box 440, Boulder, CO
80309-0440}

\begin{abstract}
  The broad X-ray iron line seen in the spectra of many AGN is thought
  to originate from the inner regions of the putative black hole
  accretion disk, and hence provides a rare probe of that central
  region.  In principle, future high throughput X-ray spectrometers
  should be able to examine the temporal response of this fluorescent
  line to flares in the X-ray continuum source (which energizes this
  emission line) --- i.e. iron line reverberation mapping will be
  possible.  It has been previously found that there are robust
  reverberation signatures of the black hole spin, mass and the X-ray
  flare location above the accretion disk.  Here, we simulate
  observations of a bright Seyfert nucleus with the proposed NASA
  mission {\it Constellation-X} and demonstrate the feasibility of
  detecting these reverberation signatures with this mission.  Hence,
  starting with {\it XMM} in 1999, and maturing with {\it
  Constellation-X} in c2010, iron line reverberation will open a new
  window on the innermost regions of AGN.
\end{abstract}

\keywords{accretion, accretion disks --- black hole physics --- galaxies:
active --- galaxies: Seyfert --- X-rays: galaxies}

\section{Introduction}

X-ray spectroscopic observations of Active Galactic Nuclei (AGN) have
provided the first direct probe of the innermost regions of black hole
accretion disks which are subject to strong-field general relativistic
effects. X-ray illumination of the surface layers of the accretion
disk produces a strong iron K$\alpha$ fluorescence line (George \&
Fabian 1991; Matt, Perola \& Piro 1991) which is broadened and skewed
in a characteristic manner by the Doppler motion of the disk and
gravitational redshifting (Fabian et al. 1989; Laor
1991). Observations of Seyfert 1 galaxies by the Advanced Satellite
for Cosmology and Astrophysics (\emph{ASCA}) have confirmed the
presence of such lines with the expected broad and skewed profiles
(Tanaka et al. 1995; Nandra et al. 1997). Alternative mechanisms
(i.e. those not involving black hole accretion disks) for the
production of such a broad and skewed line profile appear to fail
(Fabian et al. 1995).

The X-ray emission from a typical AGN shows significant short
timescale variability, with large amplitude changes in flux in
hundreds of seconds. This variability is likely to be associated with
the activation of new X-ray emitting regions either above the
accretion disk or in a disk-hugging corona. These flares produce an
`echo' in the form of a fluorescent iron line response from the disk.
The line profile depends upon the space-time geometry, the accretion
disk structure and the pattern of the X-ray illumination. Given the
low count rate of typical AGN in \emph{ASCA}, it is necessary to
observe for long periods ($\sim 10^4$ secs) to obtain sufficiently
high signal-to-noise to study the Fe K$\alpha$ emission line
profile. Such long exposure times include several hundred light
crossing times of a gravitational radius ($1 GM/c^3=500$ secs for a
$10^8M_\odot$ black hole) so that the observed line profile is
\emph{time-averaged} and details of any variability are lost, and this
lack of a timescale does not allow the black hole mass to be
determined. The time-averaged line profiles possess a degeneracy that
does not allow the spin parameter of the black hole to be simply
determined; extremely different space-time geometries and illumination
patterns may produce almost identical time-averaged line profiles
(Reynolds \& Begelman 1997), although their absorption effects differ
(Young, Ross \& Fabian 1998). Detailed time-resolved observations of
such emission line variability using the technique of
\emph{reverberation mapping} (Blandford \& McKee 1982; Stella 1990;
Reynolds et al.  1999, hereafter R99) allows the degeneracies
mentioned above to be broken. Reverberation mapping has already been
used with great success in the optical and UV wavebands, eg. the AGN
Watch results on NGC~5548 (Peterson et al. 1999; AGN Watch). R99 have
calculated the iron line response as seen by distant observers for a
number of scenarios. Their method is outlined below.

The activation of a new X-ray flaring region is approximated by an
instantaneous flash from an isotropic point source above the accretion
disk, whose location is specified in Boyer-Lindquist coordinates. The
accretion disk is assumed to be geometrically thin and confined to the
equatorial plane. Whilst the disk may extend out to large radii we are
only interested in the line emitting region within $\sim 50$
Schwarzschild radii ($50 R_{\rm s}=100 GM/c^2$). The disk is divided
into two regions, that inside the radius of marginal stability,
$r_{\rm ms}$, where there are no stable circular orbits and material
plunges into the black hole, and that outside $r_{\rm ms}$ where
material is assumed to follow essentially Keplerian orbits. $r_{\rm
ms}$ is a decreasing function of $a$, the dimensionless angular
momentum per unit mass of the black hole, decreasing from $6 GM/c^2$
for a Schwarzschild black hole ($a=0$) to $\sim 1.2 GM/c^2$ for a
maximally spinning Kerr black hole ($a=0.998$), assuming that the
accretion disk is in a prograde orbit.

The ionization state of the accreting material is determined by
considering the ionization parameter $\xi=4\pi F_{\rm x}/n_{\rm e}$,
where $F_{\rm x}$ is the illuminating X-ray flux and $n_{\rm e}$ is
the electron density. Outside $r_{\rm ms}$ the electron density is
very large and the disk is `cold', with iron less ionized that Fe
XVII. Inside $r_{\rm ms}$ the density drops rapidly as material
plunges into the black hole and, if this region is illuminated, it may
become photoionized, which affects the strength of the fluorescence
line that is produced. We use the following prescription

\begin{enumerate}
\item $\xi<100$ --- cold fluorescence line at 6.4 keV
\item $100<\xi<500$ --- no line emission due to resonant trapping and Auger
  destruction of line photons
\item $500<\xi<5000$ --- a combination of He-like and H-like lines at
  6.67 keV and 6.97 keV each with an effective fluorescent yield equal
  to that for the neutral case
\item $\xi>5000$ --- no line emission since material is completely ionized.
\end{enumerate}

The X-ray efficiency of the source $\eta_{\rm x}$ is defined as
$\eta_{\rm x}=L_{\rm x}/\dot{m}$, where $L_{\rm x}$ is the X-ray
luminosity and $\dot{m}$ is the mass accretion rate. In general, the
higher the source efficiency the more highly ionized the material
within $r_{\rm ms}$ becomes.  This simple model of the accretion disk
is sufficient for our purposes.

Photon paths are traced from the flare to compute the illumination
pattern on the disk as a function of time, and the corresponding iron
line response. The evolution of the line profile as seen by an
observer located at a given inclination to the accretion disk is then
calculated. The line response to an instantaneous ($\delta$-function)
flare is often referred to as the \emph{transfer function}. For
further details on the calculation of these transfer functions the
reader is referred to R99.

R99 note that transfer functions contain a number of robust indicators
of the space-time geometry and the location of the X-ray emitting
regions. It is possible to determine the location of the flares,
whether they be on or above the accretion disk, on the approaching or
receding side of the disk, or along its rotation axis. It is also
possible to differentiate between Schwarzschild and maximally spinning
Kerr black holes.

In this paper we simulate Constellation X-ray Mission
(\emph{Constellation-X}) observations of the iron line response to
such X-ray flares using the transfer functions of R99.
\emph{Constellation-X} will be the first X-ray observatory
with sufficient sensitivity to detect a significant Fe K$\alpha$ line
within a light crossing time for nearby bright Seyfert 1 galaxies (the
proposed effective area of \emph{Constellation-X} around the iron line
energy is about 5 times that of the X-ray Multi-Mirror satellite
(\emph{XMM}) which in turn is about 20 times that of \emph{ASCA}).

The aim of this work is to assess the feasibility of detecting the
various reverberation signatures described in R99 which will act as
probes of the black hole spin and mass.  We expand upon an issue
raised by R99 and show that a `red-ward moving bump' in the iron line
profile is a robust signature of a black hole with spin parameter
$a>0.9$.  We also step beyond the single $\delta$-function flare case
and examine whether reverberation from realistic multiple flare cases
can be disentangled using an instrument such as
\emph{Constellation-X}.   

\section{Simulation method}

Our simulations are designed to represent the case of a
\emph{Constellation-X} observation of an X-ray bright Seyfert 1 galaxy such
as MCG--6-30-15 or NGC~3516.  These sources show considerable X-ray
continuum flux variability on all timescales, with evidence for flares
on short timescales (see Fig.~1).  Taking parameters for MCG--6-30-15,
a typical average continuum flux in the range 2--10 keV is $6 \times
10^{-3}$ ph s$^{-1}$ cm$^{-2}$.  The average equivalent width of the
fluorescent iron line is 300~eV (Tanaka et al. 1995) which corresponds
to a 2--10 keV flux of $10^{-4}$\,ph\,s$^{-1}$\,cm$^{-2}$. These
figures may be used to estimate the line flux expected for a given
continuum flux, assuming the efficiency of conversion of continuum to
line photons remains unchanged.

\begin{figure*}\centerline{\psfig{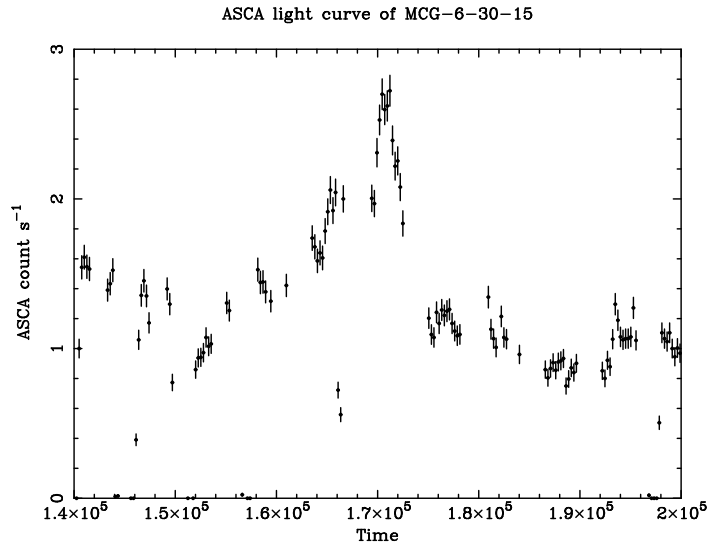}}
\caption{60000 second ASCA light curve of MCG--6-30-15 showing two `flares'
  lasting a few thousand seconds each (Lee et al 1999).}\end{figure*}

Initially, we assume a continuum level comparable to the average flux
of NGC~3516 ($1.2\times10^{-2}$ ph s$^{-1}$ cm$^{-2}$) which is
constant apart from an instantaneous ($\delta$-function) flare at some
time and localized to a point source above the disk.  The flare is
assumed to have the equivalent of 10000 seconds of continuum flux.
This corresponds to a flare lasting 1000 seconds with 10 times the
continuum flux or a flare lasting 5000 seconds with twice the
continuum flux. The duration of the flare does not significantly
influence our results as long as it is relatively short lived, lasting
only a few $GM/c^3$. Longer flares would result in the transfer
functions being blurred over time; a possibility we discuss in section
3.5. We do not specify a particular spectrum for the flare but assume
that the fraction of flux from the flare that is converted into line
photons is the same as that inferred from the time-averaged
spectrum. \emph{Constellation-X} observations of this system,
including photon counting statistics, are then simulated using current
estimates of this observatory's (energy dependent) effective
area\footnote{Effective area curves were obtained from the NASA
Goddard Space Flight Center web-page:
http://constellation.gsfc.nasa.gov/www/area.html}. The line profile
was added to a power-law continuum of photon index 2 which was
subtracted from the overall simulated data in order to yield a
simulated, time-varying, iron line profile.

The mass of the black hole determines the timescale on which
reverberation effects occur. For a $10^8\Msun$ black hole, for
example, the light crossing time of one gravitational radius is $1
GM/c^3=500$\,sec. This is the time period over which we simulate and
record individual iron line profiles in order to study the
observability of various reverberation effects. For higher mass black
holes the reverberation signatures would be more readily observed
since longer integration times may be used. The converse is true of
lower mass black holes.

\section{Results}

\subsection{Spin parameter}

Measuring black hole spin (and, implicitly, testing general relativity
in the strong field limit) is one of the main motivations for studying
iron line reverberation. If an X-ray flare occurs along the symmetry
axis above the accretion disk, the light echo will split into two
distinct `rings', one of which propagates outwards to large radii
simply due to light travel time, and one which propagates
asymptotically towards the horizon. This last feature is due to the
progressively more severe relativistic time delays suffered by photons
passing close to the black hole. It produces a red-bump in the line
profile which moves to lower energies as time goes on.  Equivalently,
it produces a `red-wing' in the transfer function. Fig.~2 shows
simulated transfer functions for values of the spin parameter $a$
between 0--0.99. For these calculations a low source efficiency
$\eta_{\rm x}=10^{-3}$ was chosen to maximize the region within
$r_{ms}$ that may produce highly redshifted fluorescent line
emission. A more realistic, higher value of $\eta_{\rm x}$ would
result in the region within $r_{\rm ms}$ becoming more highly ionized
and less able to produce fluorescent line emission (Reynolds \&
Begelman 1997). The precise shape of the `red-wing' depends upon the
spin parameter.  This red-wing only has a pronounced slope in the case
of near-extremal Kerr black holes ($a>0.9$).  Only in these cases will
the corresponding line profile possess a red-bump which moves to lower
energies as time progresses.  We conclude that this feature is a
robust signature of near-extremal Kerr black holes.

\begin{figure*}\centerline{\psfig{figure=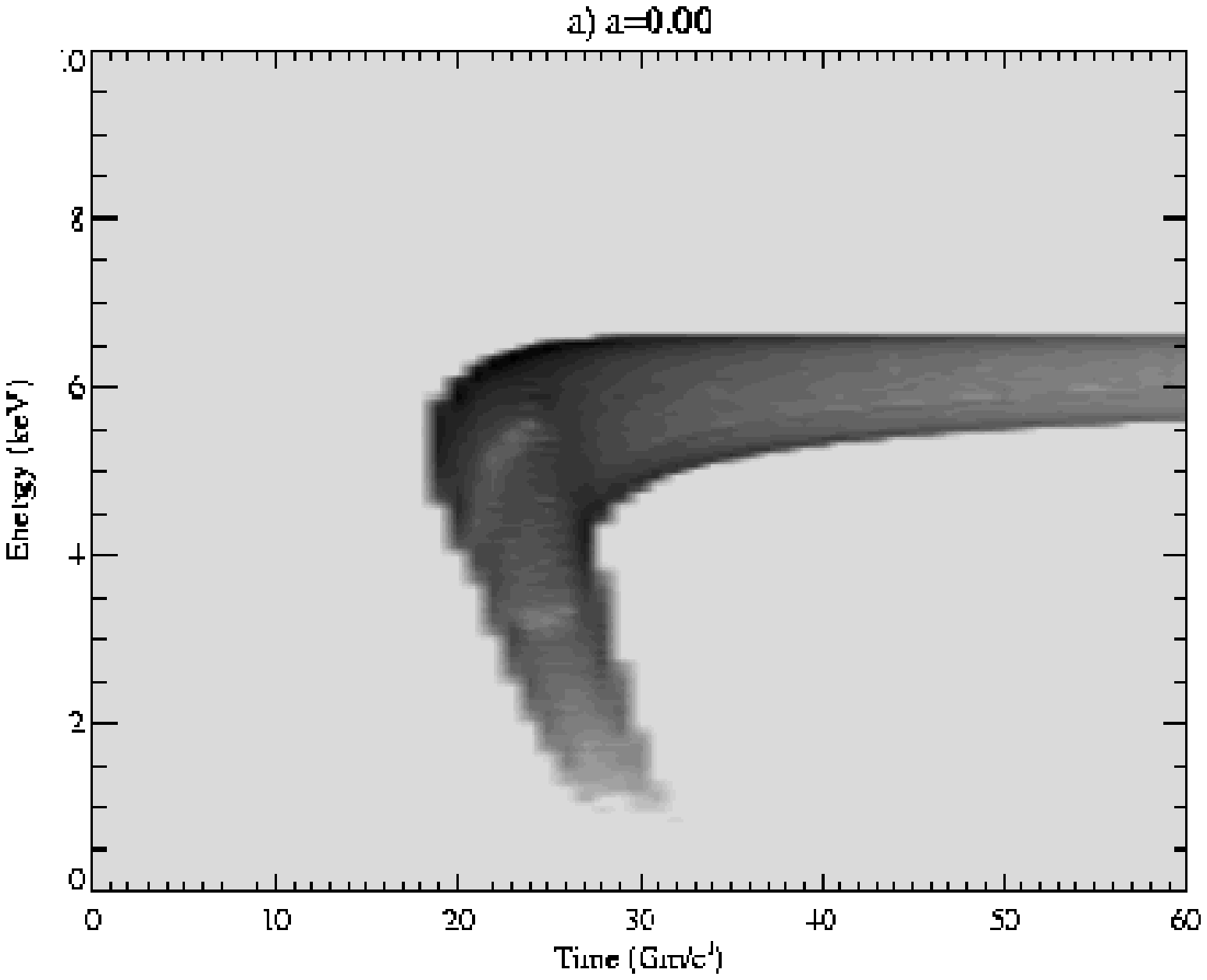,width=0.3\textwidth}\psfig{figure=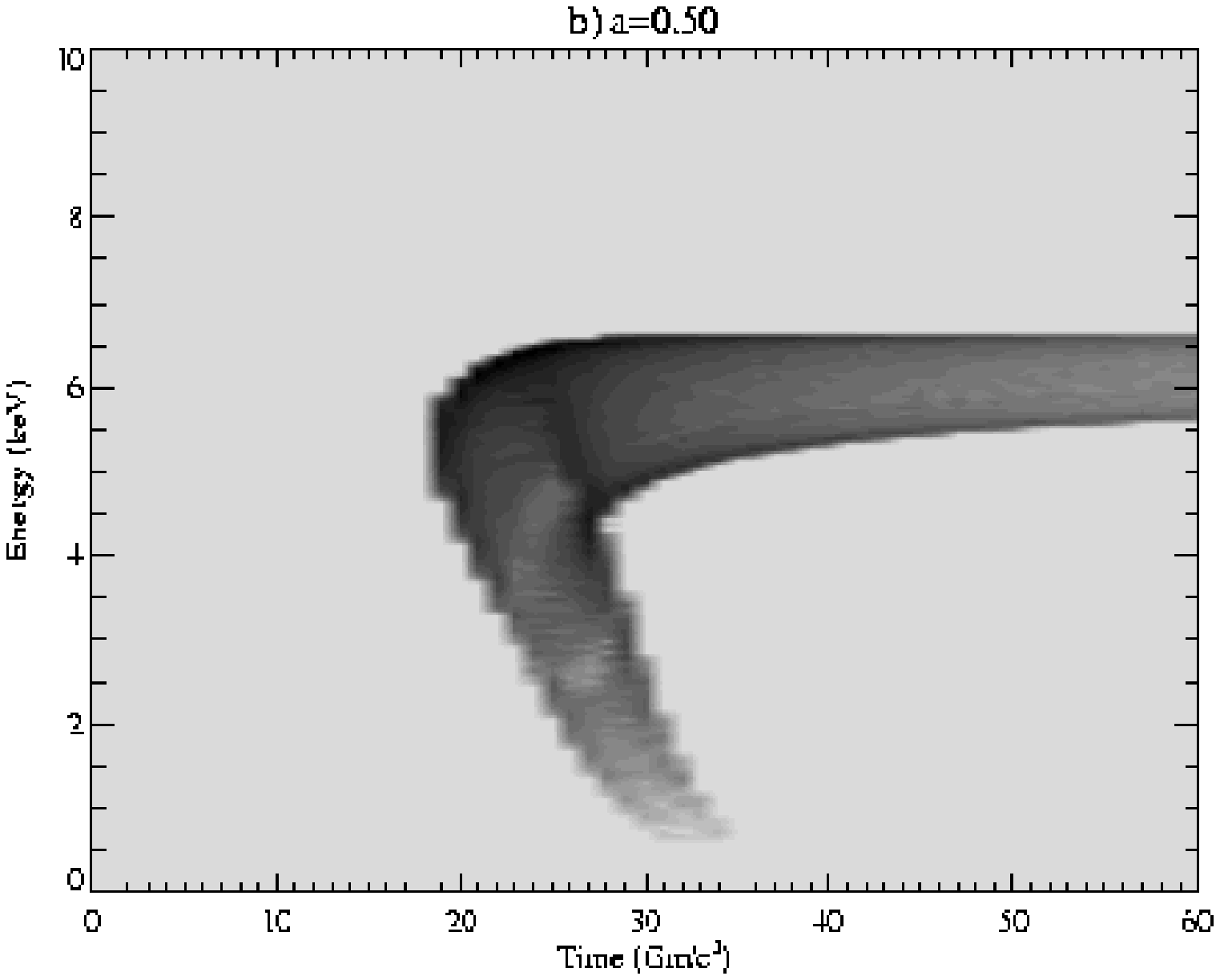,width=0.3\textwidth}\psfig{figure=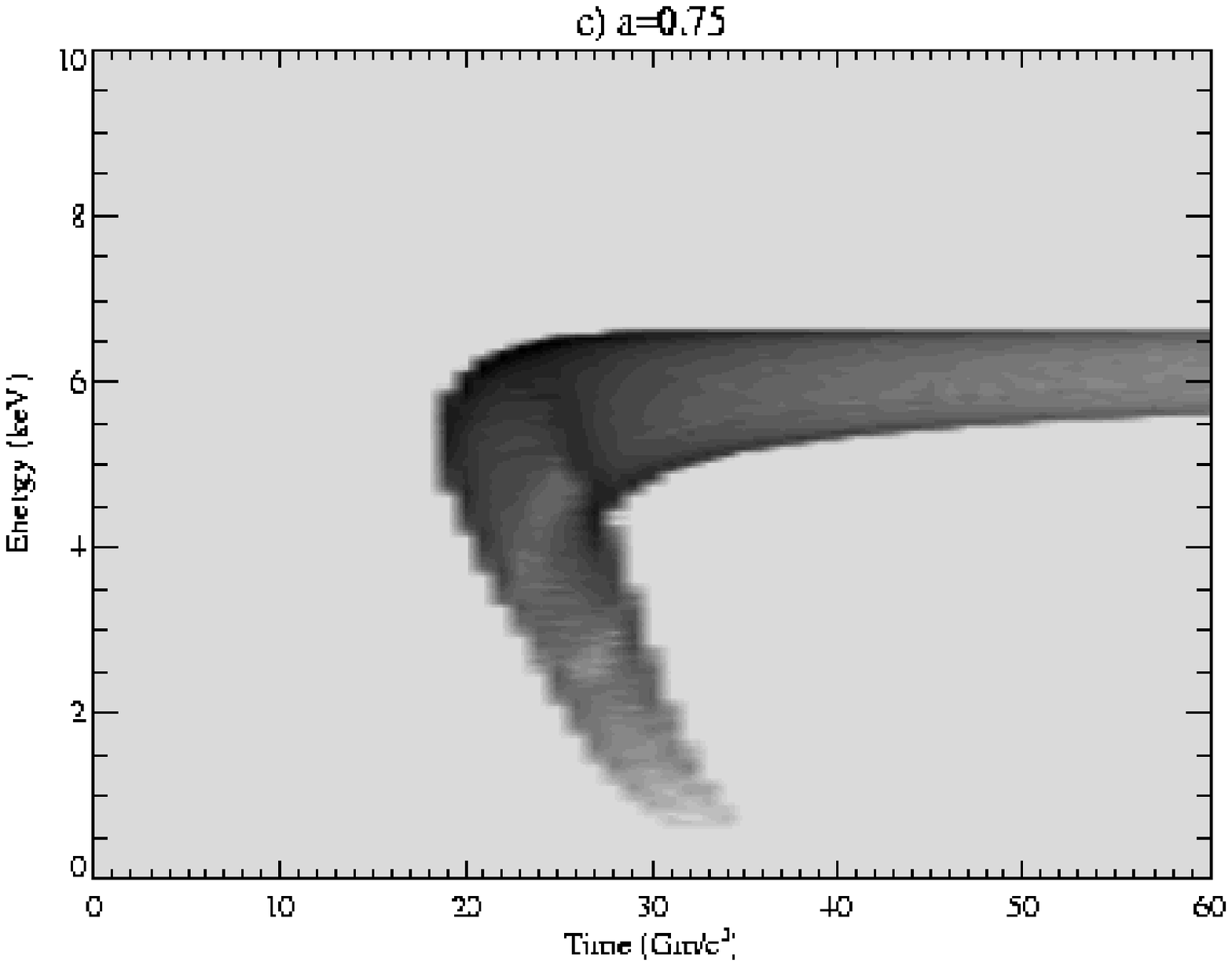,width=0.3\textwidth}}\centerline{\psfig{figure=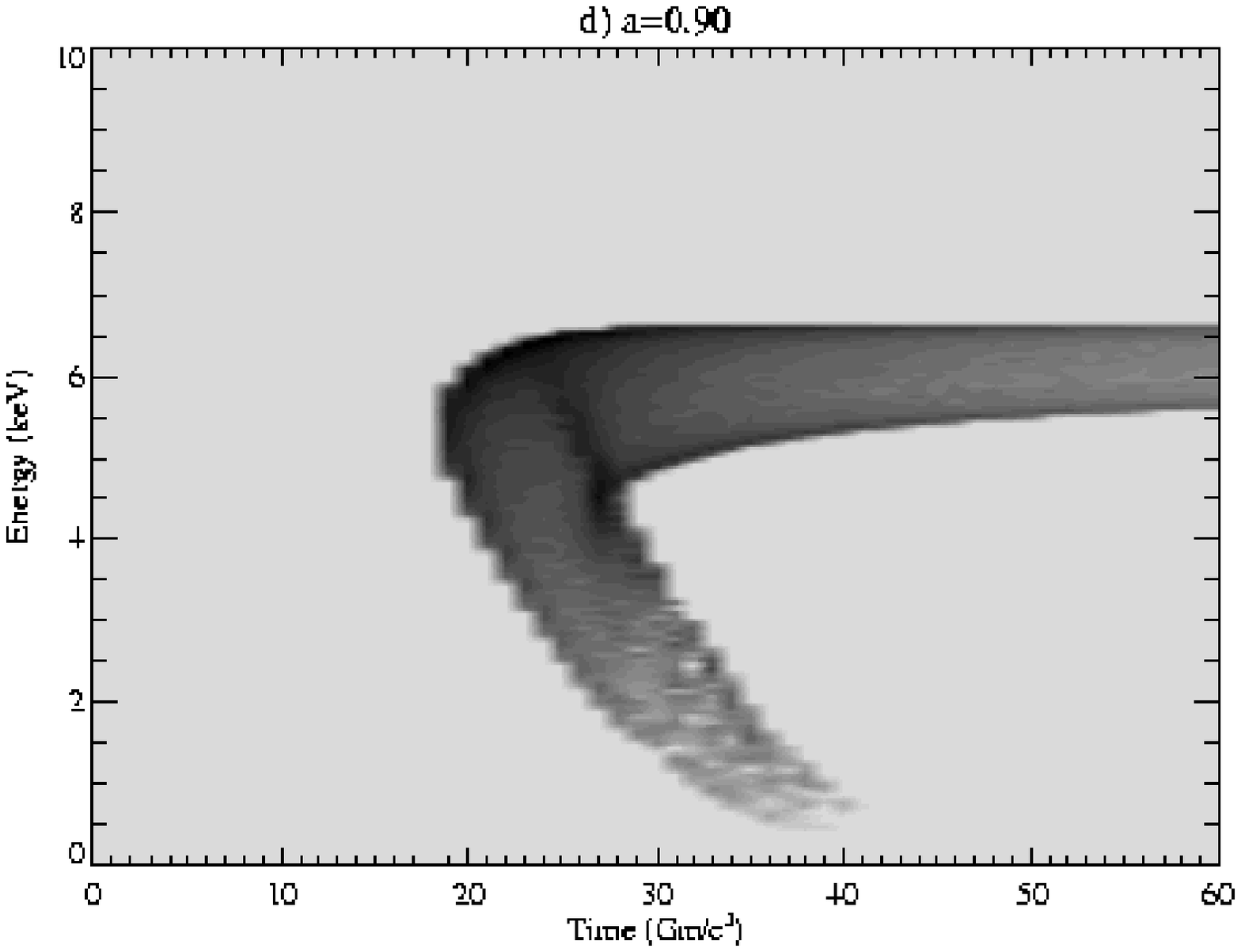,width=0.3\textwidth}\psfig{figure=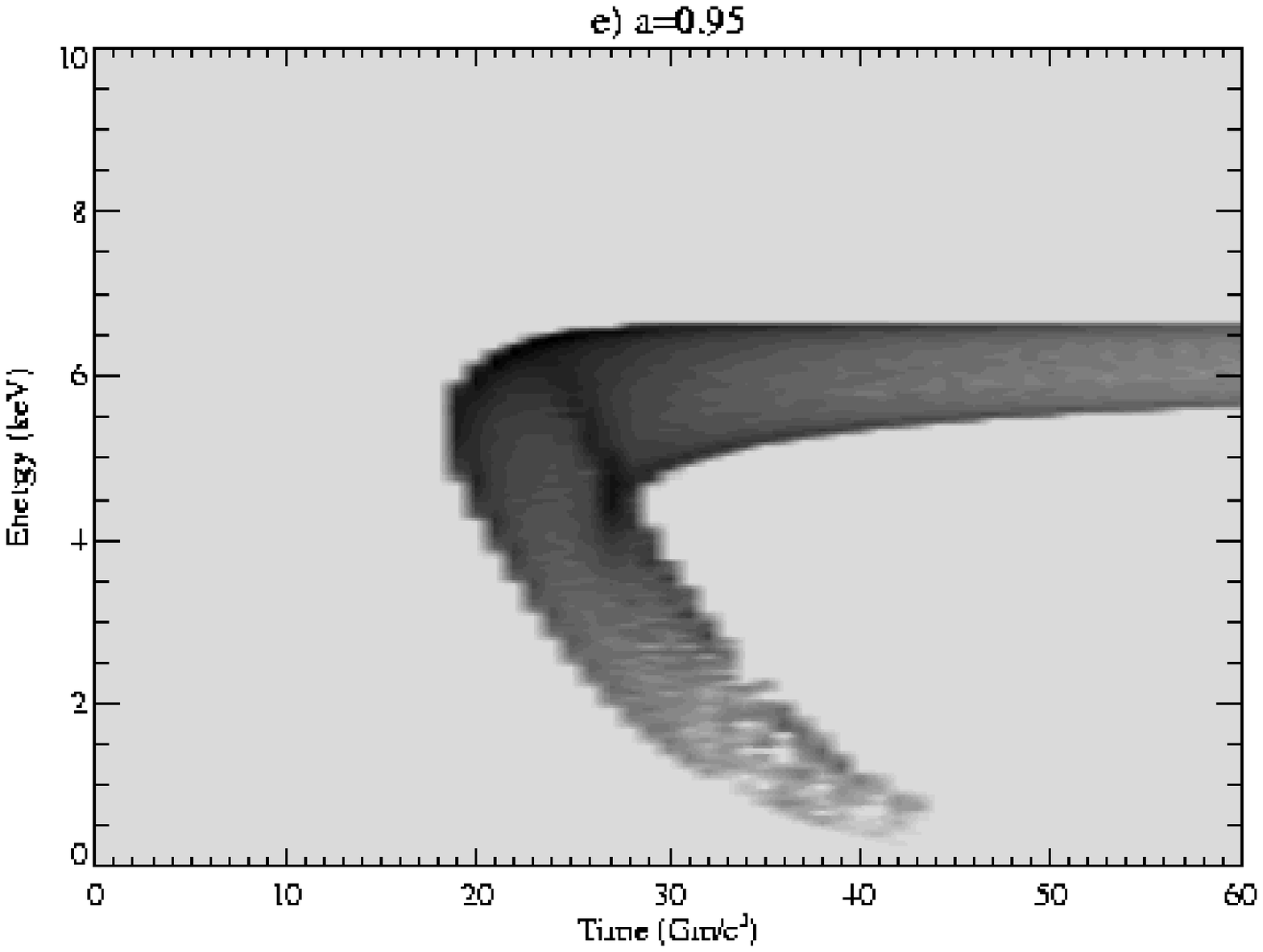,width=0.3\textwidth}\psfig{figure=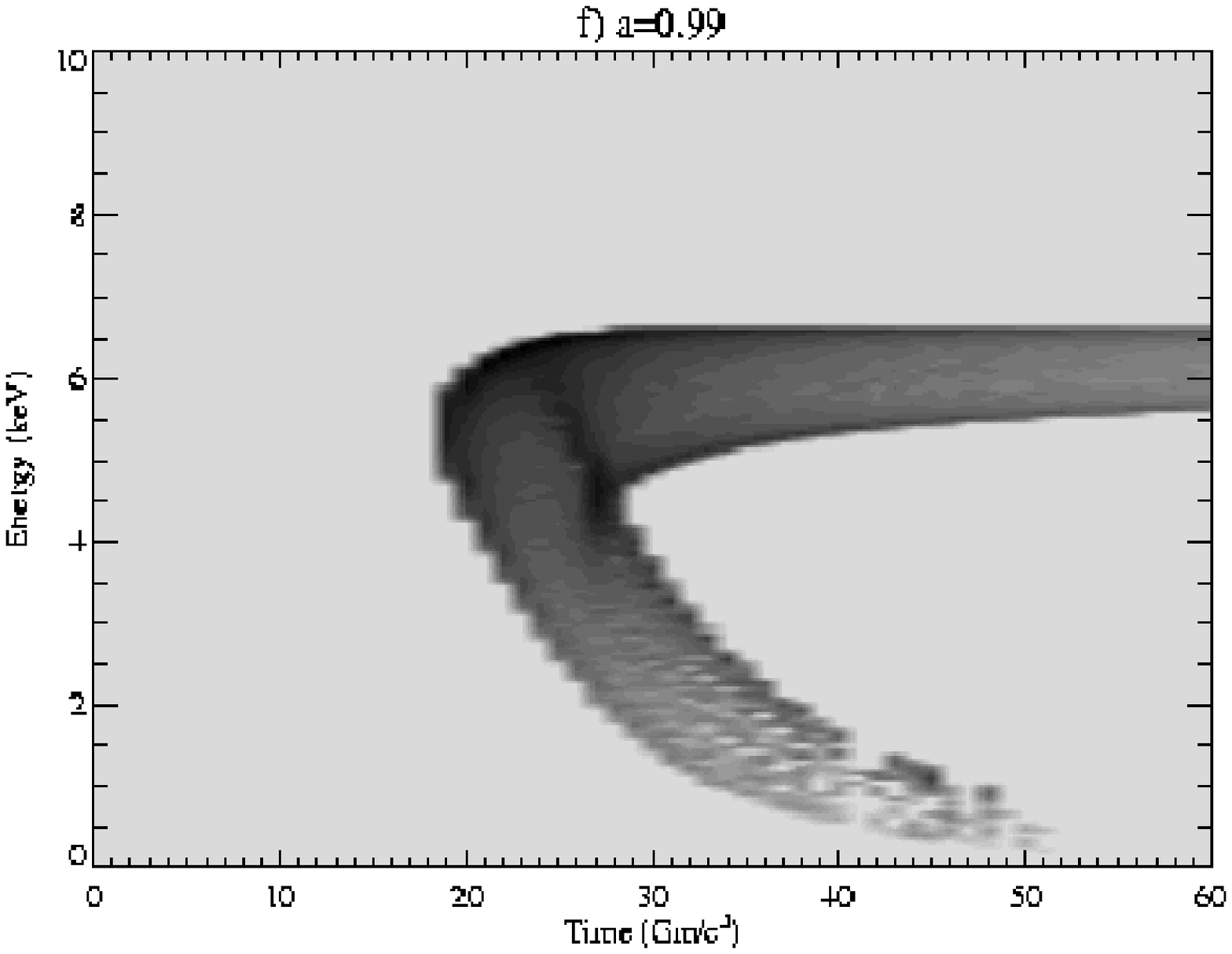,width=0.3\textwidth}}
\caption{Theoretical transfer functions of Fe K$\alpha$ line for different
  spin parameters, (a) $a=0$, (b) $a=0.5$, (c) $a=0.75$, (d) $a=0.9$,
  (e) $a=0.95$ and (f) $a=0.99$. The extreme `red-tail' is a robust
  signature of a rapidly spinning black hole. The flare has been
  placed on the symmetry axis at a height of $10GM/c^2$ above the disk
  plane, with an observer inclination of $30\degmark$. The source
  efficiency is $\eta_{\rm x}=10^{-3}$.}\end{figure*}

\begin{figure*}
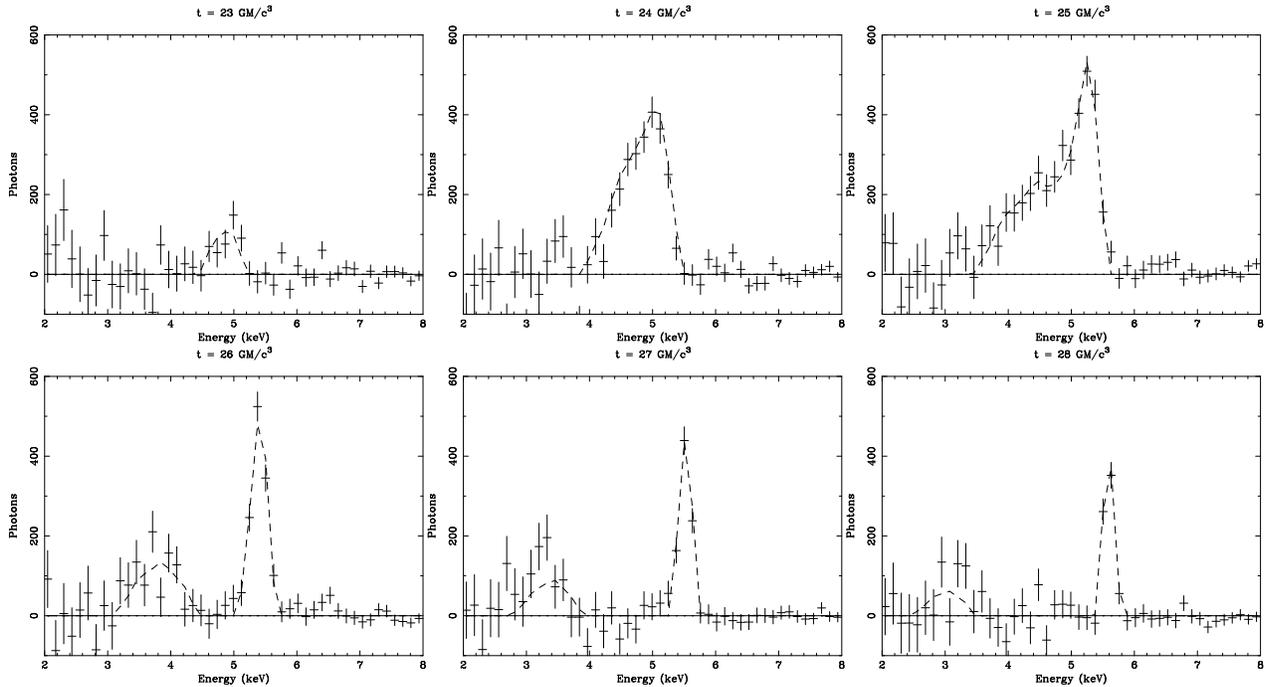
\centerline{\psfig{figure=f3a.ps,width=0.3\textwidth,angle=270}\psfig{figure=f3b.ps,width=0.3\textwidth,angle=270}\psfig{figure=f3c.ps,width=0.3\textwidth,angle=270}}\centerline{\psfig{figure=f3d.ps,width=0.3\textwidth,angle=270}\psfig{figure=f3e.ps,width=0.3\textwidth,angle=270}\psfig{figure=f3f.ps,width=0.3\textwidth,angle=270}}
\caption{$2GM/c^3$ (1000\,s) simulated observations of the red-ward moving
  bump between times $23 GM/c^3$ and $28 GM/c^3$ for an almost face-on
  (inclination $3\degmark$) accretion disk. The flare occurred on the
  rotation axis of the hole/disk at a height of $10GM/c^2$ above the
  disk plane.  The red-ward moving bump is a signature of a rapidly
  spinning black hole. The dashed line shows the theoretical evolution
  of the line profile.}\end{figure*}

Fig.~3 shows the line profile in simulated $2GM/c^3$ (1000\,s)
\emph{Constellation-X} observations of the extreme Kerr case between
23--$28GM/c^3$ (11500\,s--14000\,s) after the flare is observed at
time zero. Shown here is the case of a flare on the spin axis of the
hole at a height $10GM/c^2$ above the disk plane, viewed almost
face-on, at an inclination of $3\degmark$.  The red bump is clearly
seen and its red-ward progress with time is an indicator that emission
from around a Kerr black hole is being observed. Thus, this signature
is readily observable in such a case.

\begin{figure*}\centerline{\psfig{figure=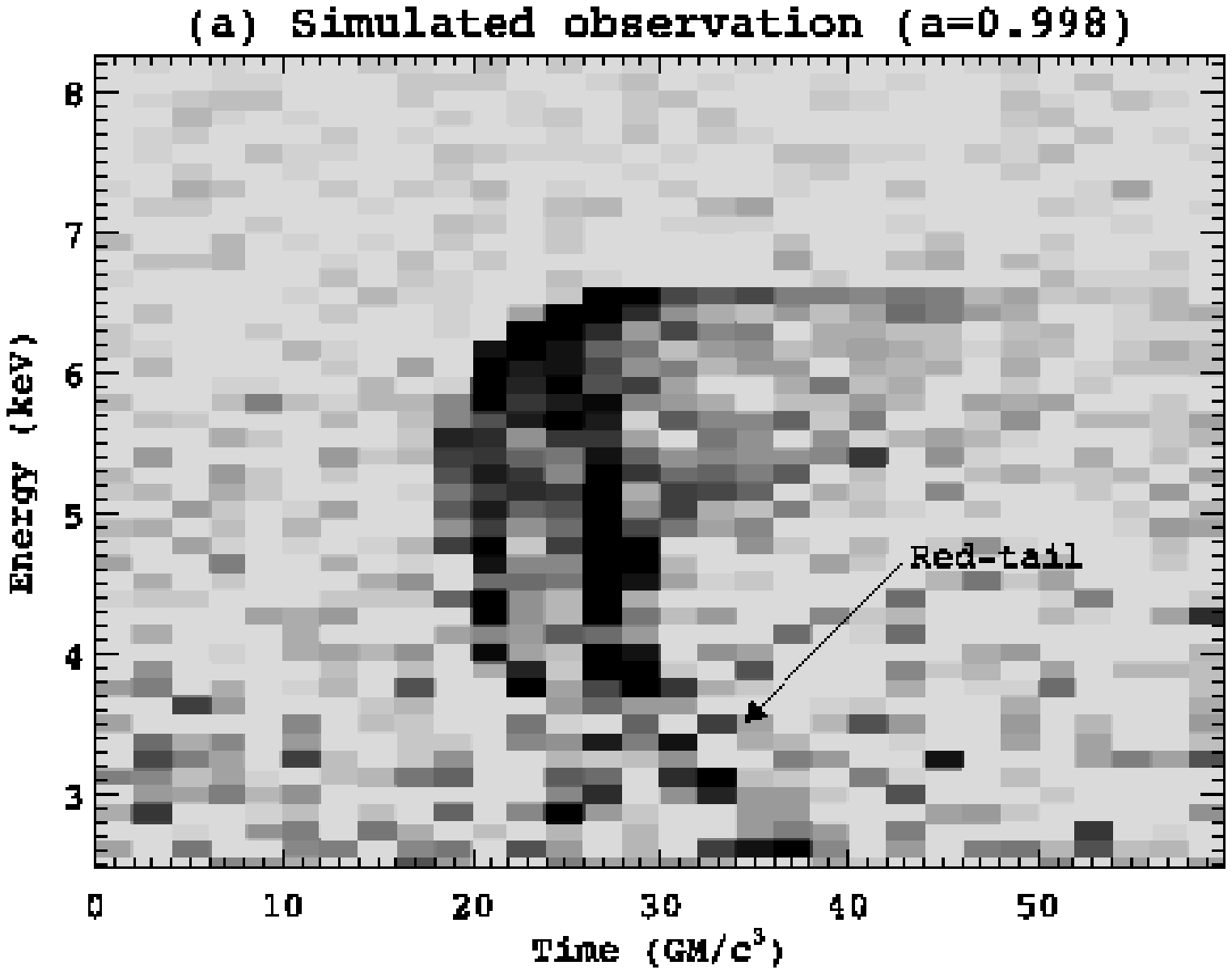,width=0.5\textwidth}\psfig{figure=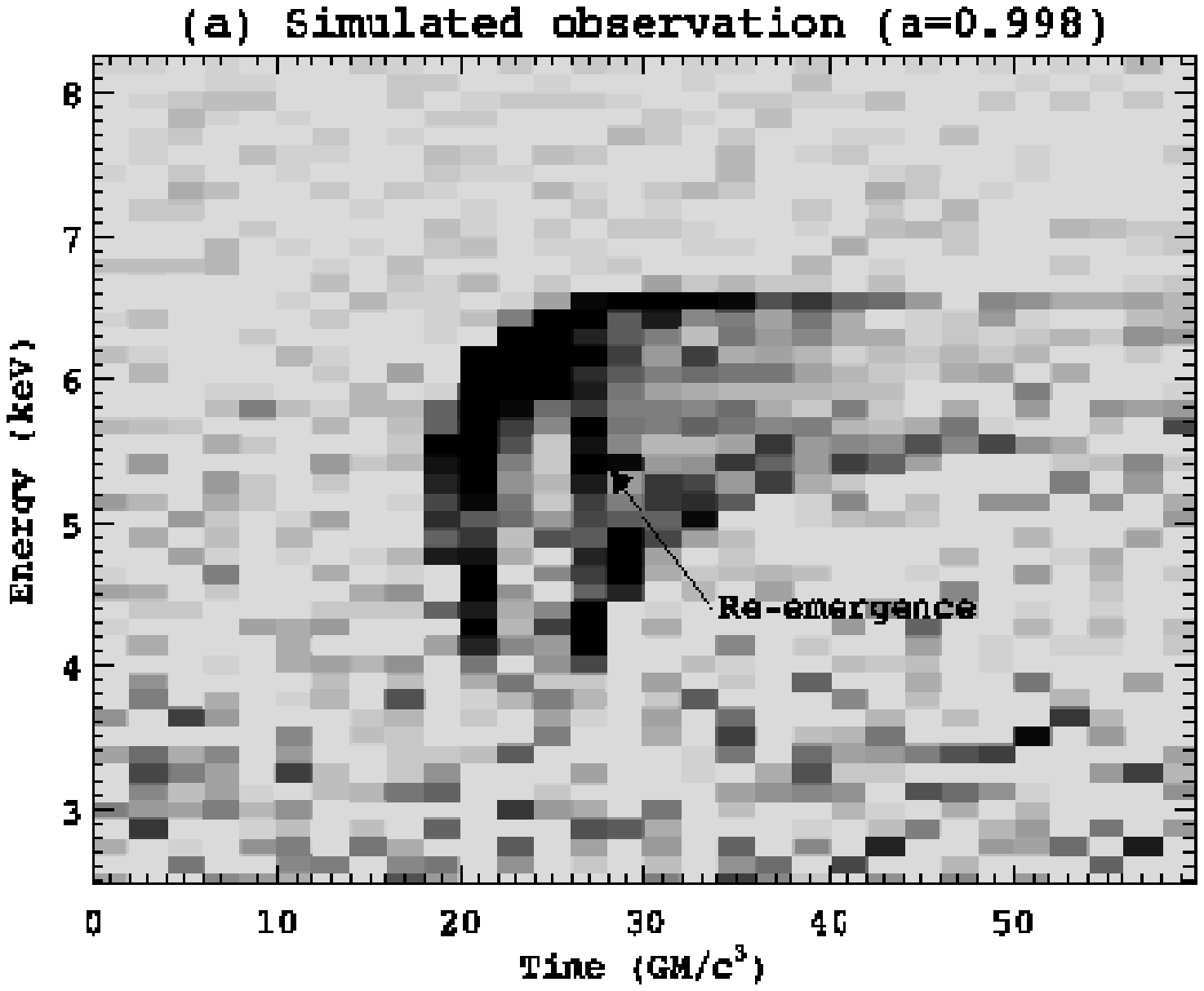,width=0.5\textwidth}}
\caption{Simulated transfer function for (a) an extremal Kerr hole, and (b)
  a Schwarzschild hole.   In both cases, the flare has been placed on
  the symmetry axis at a height of $10GM/c^2$ above the disk plane,
  and an observer inclination of $30\degmark$ has been assumed. The
  data have been rebinned to produce these figures with improved
  signal-to-noise ratio.}\end{figure*}

As the inclination of the observer increases, the photons in the
red-wing are distributed over a greater range of energies and times
and hence the detectability of the feature is reduced.  Fig.~4a shows
a simulated transfer function for a disk inclined at $30\degmark$
around an extreme Kerr black hole. The flare is assumed to be on the
symmetry axis at a height of $10GM/c^2$ above the disk plane.
Red-tail emission (at $\sim 3\keV$ and $t=30GM/c^3$) can be discerned
above the photon noise. For comparison, Fig.~4b shows the same case
but with a Schwarzschild black hole.  The differences between the
transfer functions are subtle but observable. The data in these
figures have been rebinned to increase the signal-to-noise ratio. On
the basis of these simulations, we estimate that this phenomenon is
observable for source inclinations less than $30\degmark$, although
the flare location is also an important consideration in determining
the observability of this effect.

\subsection{Confidence in the determination of the spin parameter}

The differences between the theoretical transfer functions for the
same source location but different values of the black hole spin
parameter may be seen in Fig.~2, the difference between the
Schwarzschild, $a=0$, and nearly maximally spinning Kerr, $a=0.99$,
cases being the most marked. These theoretical transfer functions may
be used as templates to fit an observation of a system inclined at
$30\degmark$ with a flare located $10 GM/c^2$ along the rotation axis
of the disk. If we simulate an observation with a particular value of
$a$ and attempt to fit that simulation with each of the template
transfer functions. Fig.~5a shows the results of fitting five
simulated observations of the case $a=0.99$. The $\Delta\chi^2$ values
are an indication of the goodness of fit (with a lower $\Delta\chi^2$
representing a better fit), normalized to $\Delta\chi^2=0$ at the best
fit values. The goodness of fit is seen to improve dramatically as the
value of $a$ used in the fit is increased indicating that the black
hole is spinning rapidly. Fig.~5b shows a similar plot for of fitting
simulated observations of the $a=0$ case. Again the $\Delta\chi^2$
values are seen to decline sharply towards the best fitting values,
and one may conclude that the black hole is not spinning rapidly. In
reality a catalogue of transfer functions would be used to fit for
different flare locations black hole spin parameters.

\begin{figure*}
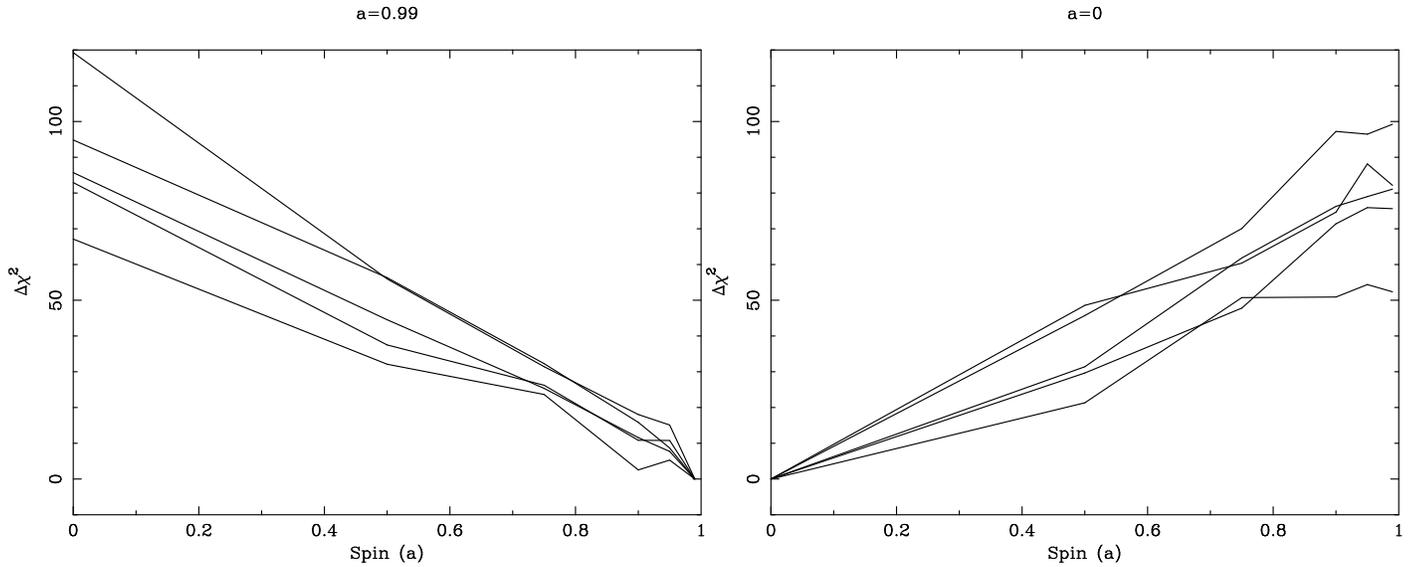
\centerline{\psfig{figure=f5a.ps,width=0.5\textwidth,angle=270}\psfig{figure=f5b.ps,width=0.5\textwidth,angle=270}}
\caption{Relative goodness of fit to simulated observations with (a)
  $a=0.99$ and (b) $a=0$. $\Delta\chi^2$ decreases with increasing
  quality of fit. Each line represents an independent
  simulation.}\end{figure*}

\subsection{Flare location and the mass of the black hole}

The location of the X-ray flares and the mass of the central black
hole (i.e. the linear dimensions of the whole system) are also
important motivations for performing iron line reverberation mapping
of AGN.  Ideally, one would take a well-measured iron line response to
a large flare and compare it to a library of theoretical transfer
functions in order to determine the location of the flare and the mass
of the hole (from the time scaling of the transfer function).  The
multi-dimensional parameter space, coupled with the limitations of
realistic data, makes this a challenging problem.

However, there are qualitative features in the transfer functions of
R99 that can be used to estimate the mass and flare location.  For
most flare locations and observer inclinations, there is a
re-emergence of the line flux (usually in the red-wing of the line) as
the observed echo of the flare works its way round to the back regions
of the disk.  The time between the initial line response and this
re-emergence is $\sim 10-20GM/c^3$.  Both the initial line response
and the re-emergence are easily observable with {\it Constellation-X}
(see Figs.~3, 4b and 6b) for $M\sim 10^8\Msun$.  Hence masses in this
range can be measured to within a factor of two or so.  Furthermore,
the time between the observed flare and the initial line response can
be compared with the above time in order to determine if the flare is
at high-latitudes above the accretion disk, or in a disk-hugging
corona.

\subsection{Observing the region within the innermost stable orbit}

\begin{figure*}\centerline{\psfig{figure=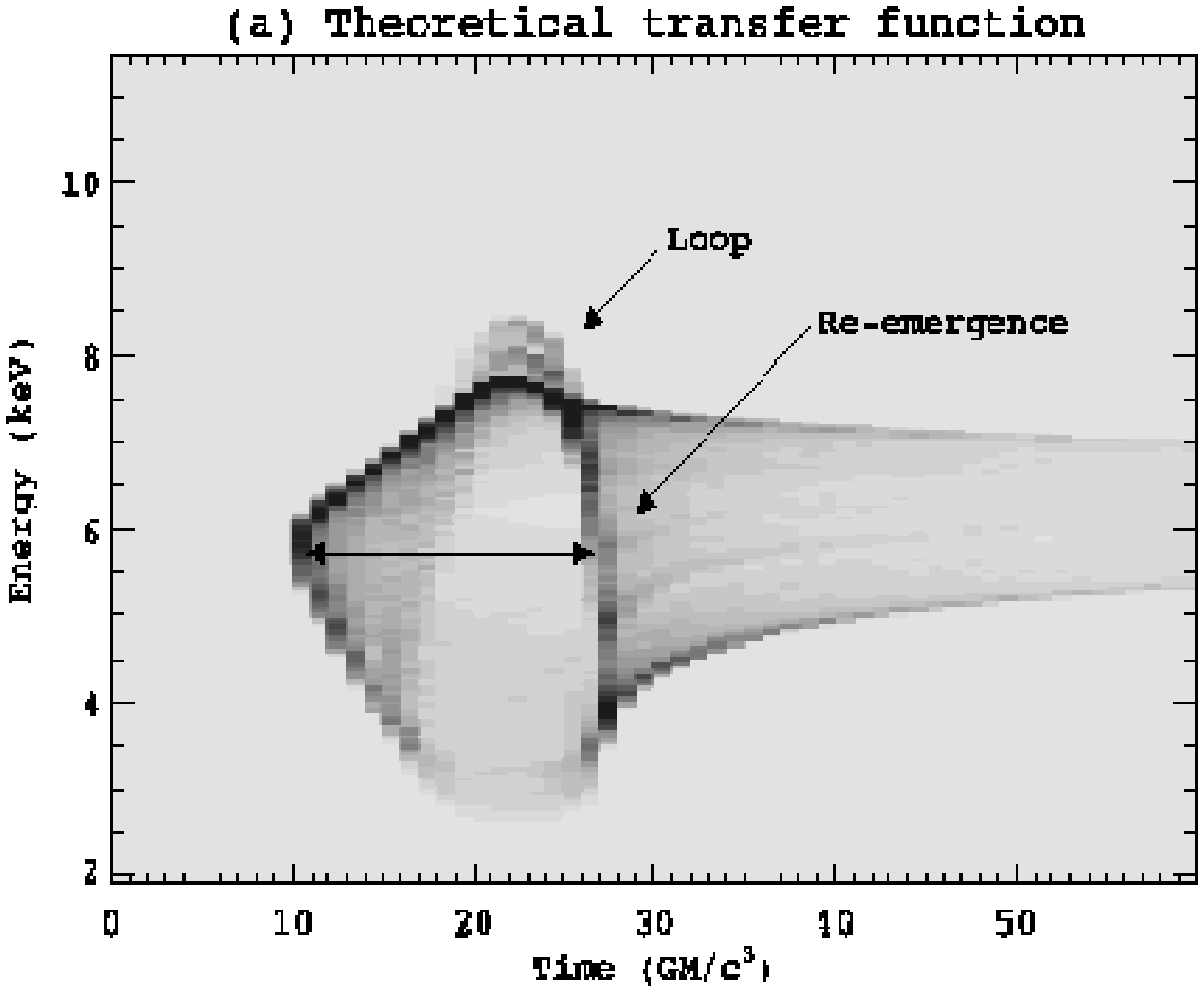,width=0.5\textwidth}\psfig{figure=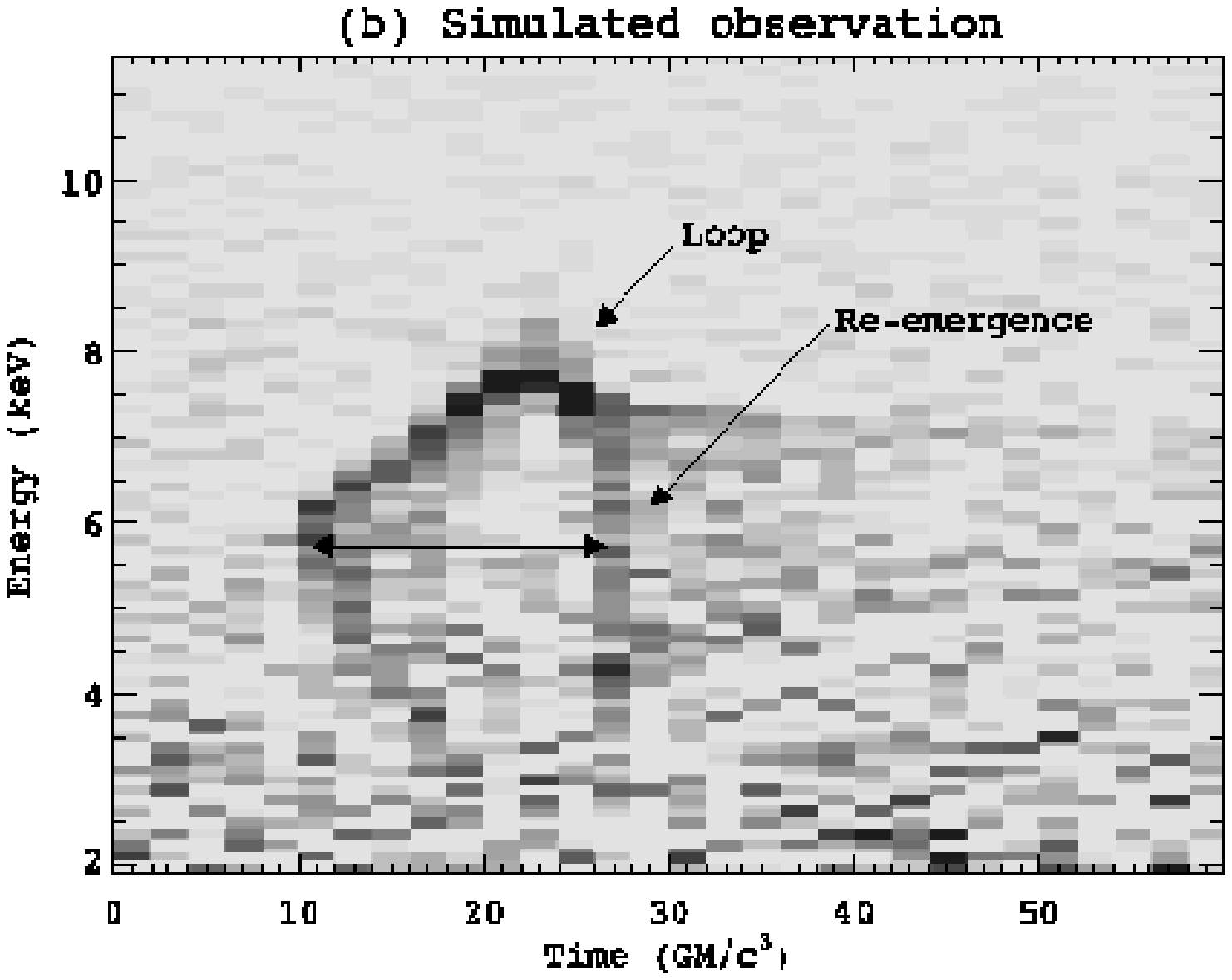,width=0.5\textwidth}}
\caption{Panel (a) shows the theoretical transfer function for a
  Schwarzschild case with an inclination of $60\degmark$ and an
  on-axis flare at a height of $10GM/c^2$. Note the `loops' in the
  transfer function corresponding to fluorescence from the ionized
  regions of the disk within the innermost stable orbit. The
  horizontal line shows the time delay between the initial response
  and the `re-emergence' which may be used to estimate the black hole
  mass. Panel (b) shows the simulated observed transfer function.
  The loops are still visible. The data have been rebinned to produce
  these figures with improved signal-to-noise ratio.}\end{figure*}

As matter crosses the innermost stable orbit its radial velocity
increases rapidly and its density drops dramatically. If this region
of the disk is illuminated by hard radiation, the matter becomes
ionized and may emit `hot' iron lines at 6.67\,keV and 6.97\,keV. In
the Schwarzschild case where the innermost stable orbit is at a radius
$6GM/c^2$ with a high latitude illuminating source, this results in a
high energy `loop' in the transfer function corresponding to the
response from this ionized region (Fig.~6; also see Fig.~1(c) and (d)
of R99). For an accretion disk at a given inclination the maximum
expected blueshift may be calculated and hence, for a `cold' iron line
at 6.4 keV, the maximum line energy can be determined.  Significant
line response at higher energies would be an indication of
fluorescence from ionized matter.  Our simulations show that these
features would be observable by \emph{Constellation-X} for inclined
disks around Schwarzschild black holes (see Figs.~6a and 6b). We have
simulated a number of observations and by fitting them with two
template transfer functions, one with the `loop' and the other
without, conclude that the presence of the `loop' is statistically
significant.  Conversely, the observation of such loops would imply
that the innermost stable orbit is some distance from the event
horizon which, in turn, would indicate a slowly-rotating black hole.

\subsection{More realistic flare models}

Considering isolated instantaneous flares is, of course, a great
simplification of the complicated activity within the
nucleus. Motivated by the light curve of Fig.~1 we consider the
slightly more realistic scenario of two overlapping flares in a source
with continuum flux comparable to that of MCG--6-30-15. Both flares
have a duration of 3000 seconds so each of their transfer functions
are smeared over $6GM/c^3$. One is on the approaching side of the disk
having the same intensity as the continuum and the other is on the
receding side of the disk with twice the intensity of the
continuum. The flare on the receding side precedes the flare on the
approaching side by $6GM/c^3$.  Fig.~7 shows the theoretical line
response to this double flare as well as the results of a
\emph{Constellation-X} simulation.  From the simulated
\emph{Constellation-X} data, one can see that there have been two flares
and can begin to disentangle the individual transfer functions. The
height of the flares above the disk along with their location should
be determinable.

It is possible that flares significantly brighter than those simulated
here may be observed and the fluorescent responses to those would be
correspondingly stronger. Other galaxies, such as NGC~3516, are twice
as bright as MCG--6-30-15 and this would help reduce the error bars
associated with the photon statistics.  Gravitational focusing of the
flare emission towards the disk (Martocchia \& Matt 1996) may further
pronounce these reverberation signatures for flares occurring
extremely close to the black hole, within $\sim 6 GM/c^2$. In such
cases the observed change in continuum flux would represent only a
fraction of the flux incident upon the disk. In addition to the bright
flares there will be a background of much smaller flares occuring
elsewhere on the disk, the response to which may be approximated by a
time-averaged line. These flares will appear as noise in the data.

\begin{figure*}\centerline{\psfig{figure=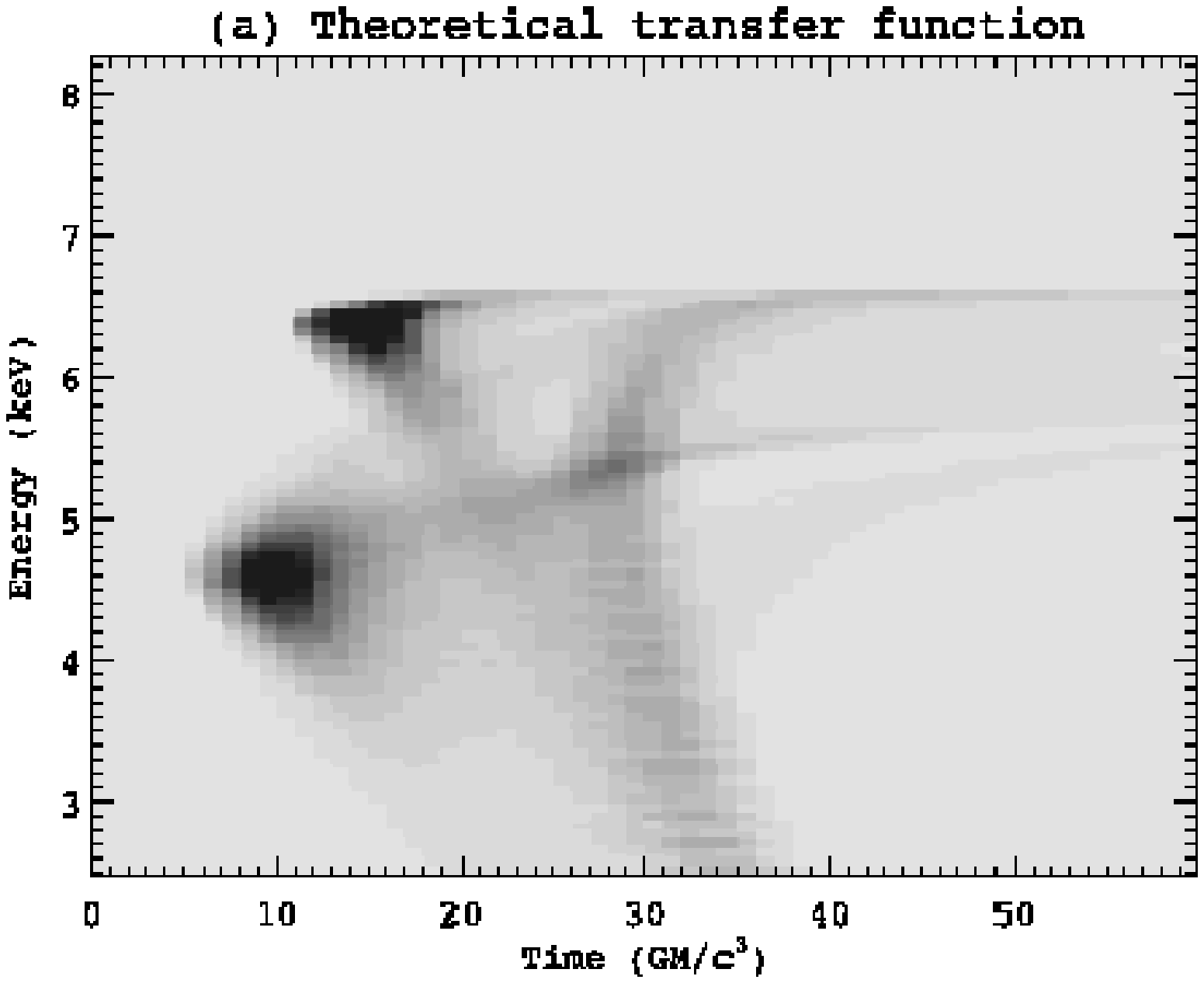,width=0.5\textwidth}\psfig{figure=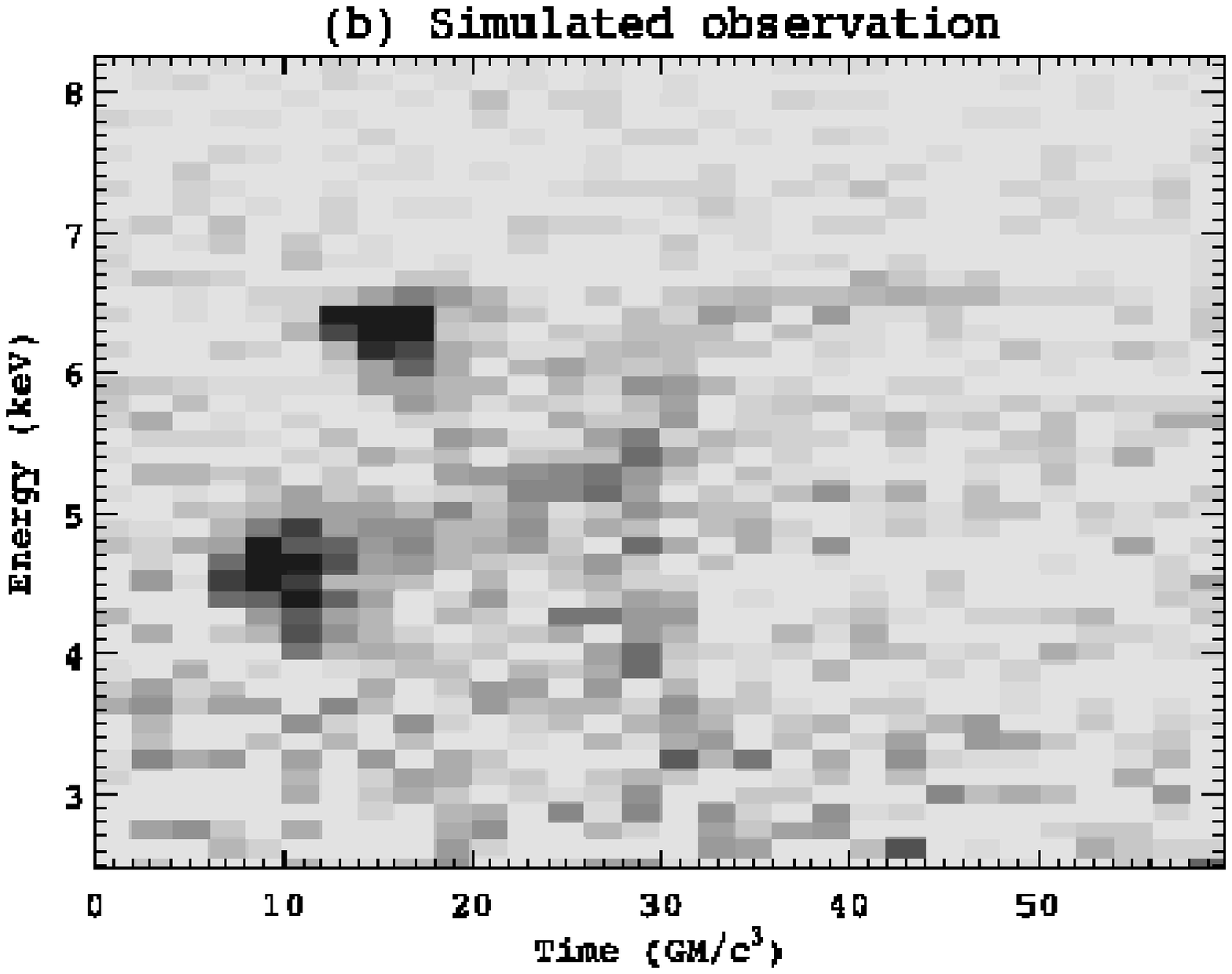,width=0.5\textwidth}}
\caption{Panel (a) shows the theoretical line response to the two
  overlapping flares described in the text.  Panel (b) shows the
  simulated line response as seen by \emph{Constellation-X}.  The
  individual transfer functions of the two flares can be
  discerned. The data have been rebinned to produce these figures with
  improved signal-to-noise ratio.}\end{figure*}

\section{Conclusions}

We have demonstrated that many of the iron line reverberation effects
noted by R99 are within reach of \emph{Constellation-X}.  In
particular, \emph{Constellation-X} will be able to search for the
red-ward moving bump in the iron line profile which is a robust and
generic signature of rapidly rotating black holes. Maximally spinning
Kerr and Schwarzschild black holes can be discriminated.  It will also
allow the time delay between a large flare and the iron line response,
as well as the form of the corresponding transfer function to be
determined.  Comparison with a library of computed transfer functions
will allow the mass of the hole and the location of the flare to be
measured.  Although this is a difficult task due to the
multi-dimensional parameter space that one must consider, we note that
there are easily determinable quantities that allow the black hole
mass and flare location to be approximated. The time delay between the
initial response and `re-emergence' of the line flux may be used to
estimate the black hole mass to within a factor of 2. The time delay
between the change in the continuum and the initial response of the
iron line, as well as the energy of this response, may be used to
estimate the location of the flare.  These studies will open up a new,
and extremely powerful, probe of the immediate environment of
supermassive black holes.

\section{Acknowledgments}

We thank Kazushi Iwasawa and Julia Lee for the unpublished recent
light curve of MCG--6-30-15, and Andy Fabian for insightful
discussion.  AJY acknowledges PPARC (UK) for support. CSR acknowledges
support from NASA under LTSA grant NAG 5-6337.  CSR also acknowledges
support from a Hubble Fellowship grant HF-01113.01-98A awarded by the
Space Telescope Institute, which is operated by the Association of
Universities for Research in Astronomy, Inc., for NASA under contract
NAS 5-26555.

\end{document}